\documentclass[twocolumn,amssymb,nobibnotes,aps,prl]{revtex4-1}
\pdfoutput=1

\usepackage{amsmath,amssymb,graphicx}
\usepackage{hyperref}

\begin{document}

\title{Discovery of Parabolic Microresonators Produced via Fiber Tapering}

\author{Dashiell L. P. Vitullo} \email{Corresponding author: vitullod+journal@gmail.com}
\affiliation{Aston Institute of Photonic Technologies, Aston University, Birmigham B4 7ET, UK}
\author{Gabriella Gardosi}
\affiliation{Aston Institute of Photonic Technologies, Aston University, Birmigham B4 7ET, UK}
\author{Sajid Zaki}
\affiliation{Aston Institute of Photonic Technologies, Aston University, Birmigham B4 7ET, UK}
\author{Kirill V. Tokmakov}
\affiliation{Aston Institute of Photonic Technologies, Aston University, Birmigham B4 7ET, UK}
\author{Michael Brodsky}
\affiliation{U.S. Army Research Laboratory, Adelphi, MD 20783-1197, USA}
\author{Misha Sumetsky}
\affiliation{Aston Institute of Photonic Technologies, Aston University, Birmigham B4 7ET, UK}






\begin{abstract}
We demonstrate a new method for creation of surface nanoscale axial photonics (SNAP) microresonators with harmonic profiles via fiber tapering in a laser-heated microfurnace. The simple procedure makes microresonators that support hundreds of axial modes with good spacing uniformity, yielding a promising prospective method for fabricating miniature frequency comb generators and dispersionless delay lines.
\end{abstract}


\maketitle

Optical microresonators with dispersionless (equidistant) spectra do not distort the temporal shape of a pulse of light stored within them \cite{Schroedinger1926,Khurgin2010,Ilchenko2003,Sumetsky2014}. This makes such optical microresonators a promising platform for dispersion-free delay lines, optical buffers, and frequency-comb generators \cite{Sumetsky2013,Sumetsky2015,Dvoyrin2016}. Microresonators with equidistant spectra over tens of axial modes have been demonstrated in miniature photonic crystal cavities \cite{Combrie2017}, semiconductor microtubes \cite{Strelow2008}, and Surface Nanoscale Axial Photonics (SNAP) \cite{Sumetsky2004, Pollinger2009}. Currently, the most promising miniature devices for frequency comb generation are toroidal microresonators which possess azimuthal eigenfrequencies with small dispersion and enable the generation of octave-spanning combs \cite{Kippenberg2011, Delhaye2007}. However, these resonators require a relatively large device footprint for generation of combs with low repetition rates \cite{Suh2018}. On the other hand, SNAP microresonators fabricated by nanoscale deformation of an optical fiber can achieve these same low repetition rates with much smaller device footprints (for microresonators with repetition rates $R_R\approx 6$ GHz: toroidal footprint $\approx 80$ mm$^2$; SNAP footprint $\approx0.06$ mm$^2$) and therefore offer a promising alternative for fabrication of miniature frequency combs which simultaneously have low repetition rate and broad bandwidth \cite{Sumetsky2013, Dvoyrin2016}. 

In this Letter, we report use of a relatively simple fiber tapering technique for fabrication of SNAP bottle microresonators with large harmonic profiles, which give rise to spectra with good eigenfrequency equidistance over hundreds of axial modes. The developed method is based on the observation made by Birks, Knight, and Dimmick in their work suggesting and demonstrating a method for high-precision characterization of optical fiber taper profiles \cite{Birks1992}. Their measurements showed that the taper waist regions, which were designed to be uniform, had small radius variation on the order of a few hundred nanometers. This variation is negligible for most applications involving light traveling through the taper, but dramatically changes the behavior of whispering gallery modes traveling around the surface of the taper, forming SNAP microresonators.

The SNAP platform leverages the exceptional uniformity of optical fiber surfaces to make optical microresonators with ultra-low loss (intrinsic quality factors $Q \approx 10^8$) \cite{Pollinger2009}. SNAP devices are usually fabricated by annealing the fiber surface with a focused $\mathrm{CO_2}$ laser, which grants sub-angstrom fabrication precision in effective radius variation (ERV) \cite{SumetskySNAPReview}. Semiparabolic SNAP bottle microresonators have been previously fabricated by scanning the laser along the fiber surface following a complicated pattern, but this process yielded low repeatability \cite{Sumetsky2013}. The newly discovered bottle microresonators made via tapering, as described below, have a small footprint ($\sim$0.06 mm$^2$), 2 mm extent, and ERV heights of order 100 nm, which is an order of magnitude larger than the height that can be produced with the standard laser annealing approach.

\begin{figure}[tb]
\begin{center}
\includegraphics[width=\columnwidth]{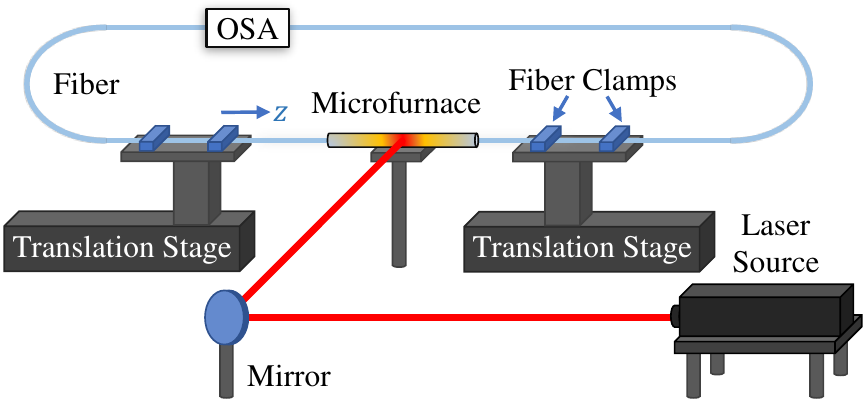}
\caption{Tapering rig illustration. A $\textrm{CO}_2$ laser heats the sapphire microfurnace. Fiber threaded through the furnace is pulled along the $z$ direction by computer-controlled translation stages. Throughout the tapering process, an optical spectrum analyzer monitors power transmission through the tapered fiber and two webcams monitor the transverse position of the fiber within the microfurnance.}
\label{fig:setup}
\end{center}
\end{figure}

\begin{figure}[htb]
\begin{center}
\includegraphics[width=\columnwidth]{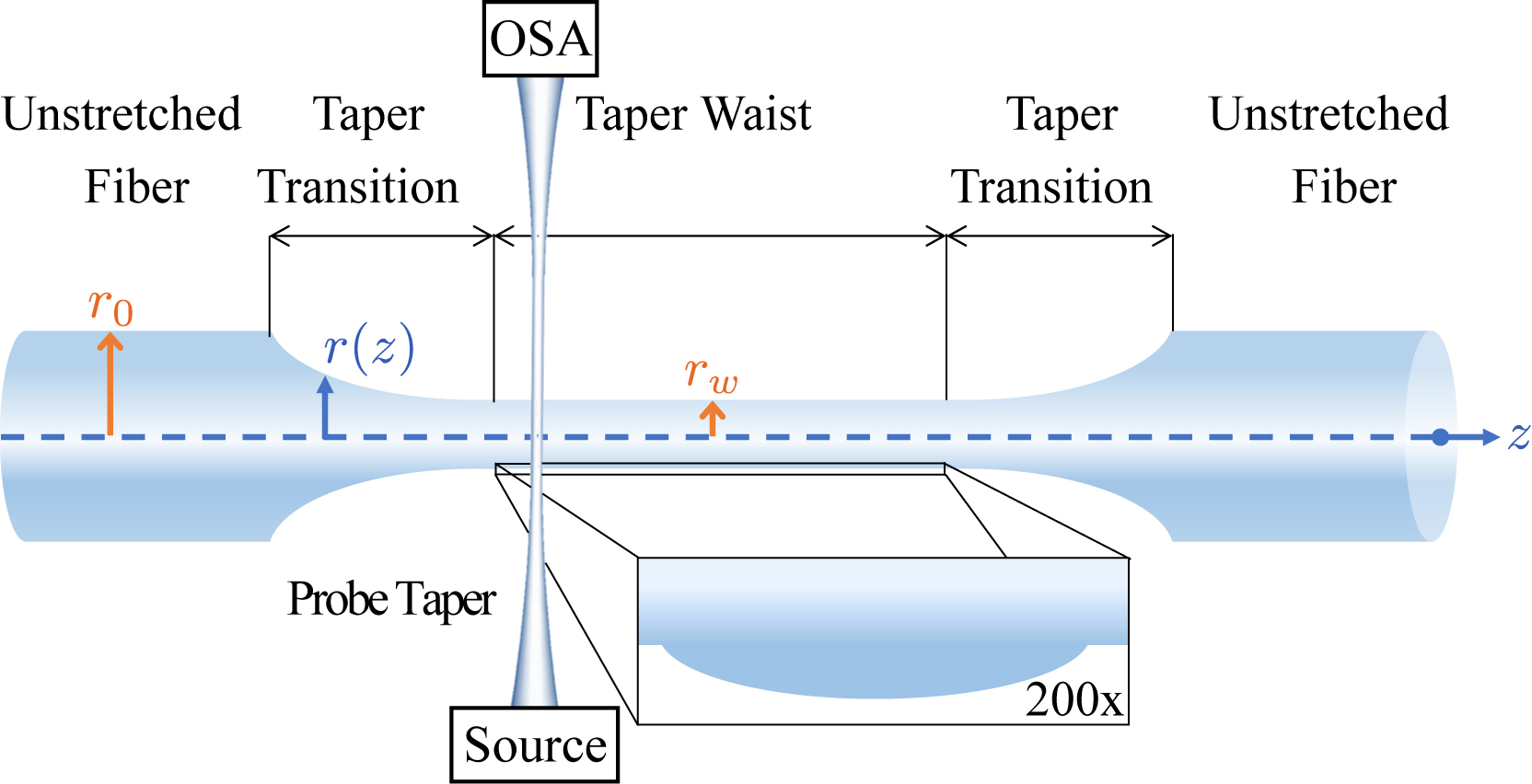}
\caption{Diagram of a tapered fiber labeled with region names and characterized with a probe taper connected to an optical spectrum analyzer (OSA) and light source. Inset shows microresonator profile.}
\label{fig:taper}
\end{center}
\end{figure}

We taper fibers with a heat-and-pull rig \cite{SumetskyMicroloop,Ward2006} where a $\mathrm{CO_2}$ laser heats a cylindrical sapphire microfurnace surrounding a 125 $\mu$m diameter optical fiber (see Fig.\ \ref{fig:setup}). The fiber is pulled with computer-controlled linear translation stages after being loaded, spliced, and positioned. The splice is positioned so it is not included in the tapered region, as its inclusion distorts the resulting SNAP device. The tapering procedure consists of moving both stages with different speeds in the same direction for some distance, completing what we call a half cycle, and then they simultaneously stop and move in the opposite direction until they have completed a full cycle. The stage with the position along the direction of motion with the largest value is referred to as the leading stage, while its partner is referred to as the trailing stage. The stages swap roles at the end of each half cycle. Let $L$ be the length of the fiber that passes through the heated portion of the microfurnace, set by the distance traveled by the trailing stage, and $dx$ be the ``stretch'' or extra distance traveled by the leading stage to pull the heated fiber. For each half-cycle of the tapering process, the trailing stage moves $L=3$ mm while the leading stage moves $L + dx = 4$ mm. This produces tapers with taper waists (i.e.\ the diameter at the thinnest point $d$) of approximately
\begin{equation}
d = d_0 \left ( \frac{L}{L+dx} \right)^{N},
\end{equation}
where $d_0=125\, \mu$m is the initial diameter of the fiber and $N$ is the number of full cycles. The tapers with harmonic microresonators described below were made using $N = 5$ cycles, and thus have a nominal waist diameter of $d=30$ $\mu$m.  We refer to one of these relatively large tapers as the taper under test (TUT). 

\begin{figure*}[htb]
\begin{center}
\includegraphics[width=\textwidth]{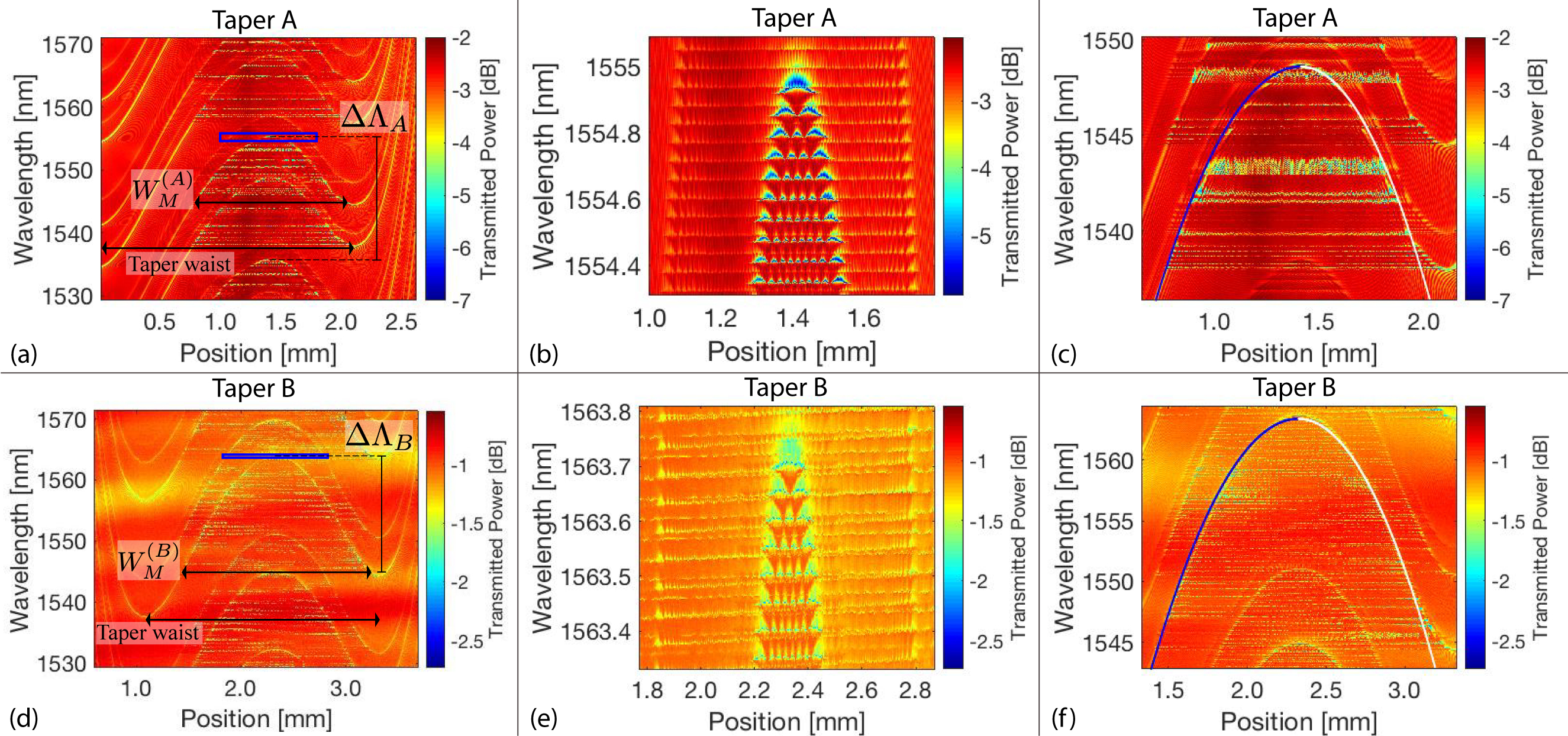}
\caption{Spectrograms of two microresonators created via tapering. Transmitted power is referenced to an input power of 0.2 mW sourced from a LUNA 5000 OVA. (a) and (d) are broadband spectrograms with multiple axial series. Spectra are measured with 2 $\mu$m separation in position along taper A and 4 $\mu$m separation along taper B. See the text for symbol definitions. The blue boxes indicate the corresponding regions shown in (b) and (e) that zoom in on the lowest order axial modes. The slight slope visible in (e) is due to temperature drift over the course of the scan. We observe different slopes for modes in different axial series in this spectrogram, and this indicates that these modes, which do not share the same values for all of $m$, $p$, and $s$, must have disparate wavelength sensitivity to temperature $\frac{\text{d}\lambda_{q,m,p,s}}{\text{d}T}$ (see \eqref{eq:resonatorLambda}). (c) and (f) show the full axial series with piecewise parabolic fits as described in the text. The fits on the right sides exclude the turn-around regions with upward concavity.}
\label{fig:spectrograms}
\end{center}
\end{figure*}

\begin{figure}[htb]
\begin{center}
\includegraphics[width=\columnwidth]{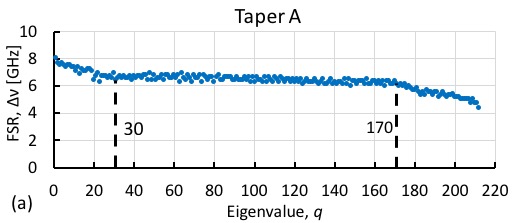}
\includegraphics[width=\columnwidth]{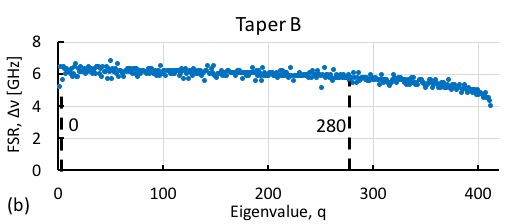}
\caption{Difference in frequency spacing between adjacent modes (FSR) vs.\ eigenvalue. More uniform spacing is observed between $q = 30$ and $q=170$ for taper A and between $q=0$ and $q=280$ for taper B.}
\label{fig:modeSpacing}
\end{center}
\end{figure}

The nominal structure of a fiber taper is diagrammed in Fig.\ \ref{fig:taper}, following the simple model for predicting taper shape given tapering parameters from \cite{Birks1992}. The actual profile of the radius $r(z)$ as a function of longitudinal position along the TUT $z$, is precisely characterized with a probe microtaper with a 1.7 $\mu$m diameter waist ($N$=15) using the method described in detail in \cite{Sumetsky2010, Kavungal2018, Sumetsky2006}. Put briefly, the probe taper is oriented transverse to the TUT and placed into contact so as to efficiently couple light propagating through the probe taper to the whispering gallery modes along the perimeter of the TUT. Interference between light transmitted past the TUT and light coupled from the whispering gallery modes back into the probe results in narrow dips in the transmission spectrum at resonant wavelengths. The radial profile $r(z)$ of the TUT can be characterized with high precision through construction of a spectrogram, which shows the transmission spectrum vs.\ position along the fiber with the transmission amplitude encoded in color (see for example Fig.\ \ref{fig:spectrograms}). The probe moves to measurement positions arranged in a uniform grid along the TUT following the path described in Section 4.2 of \cite{SumetskySNAPReview}, which prevents sticking or dragging. The whispering gallery mode structure shown in a spectrogram can be modeled using a 1D Schr\"odinger equation, and the potential is set by the effective radius variation (ERV) of the TUT.

The whispering gallery modes interrogated in a spectrogram are characterized by four quantum numbers: axial $q$, azimuthal $m$, radial $p$, and polarization $s$, and we label the resonance wavelength of a mode $\lambda_{q,m,p,s}$. The most closely-spaced modes are the series of axial modes, seen in Figs. \ref{fig:spectrograms}(b) \& \ref{fig:spectrograms}(e). The axial series is repeated at different resonant wavelengths for modes with different quantum numbers $m$, $p$, or $s$, as can be seen in Figs.\ \ref{fig:spectrograms}(a) \& \ref{fig:spectrograms}(d). The separation between adjacent azimuthal and radial values can be estimated using a semiclassical asymptotic approximation that assumes $m \gg p$, \cite{Demchenko2013,Hamidfar2018}
\begin{align}
\lambda_{m,p,s} &= \frac{2 \pi n r}{m} \left[1+\zeta_p \times \left(2 m^2 \right)^{-1/3} + \frac{n^{s}}{m\sqrt{n^2 - 1}} \right], \\
\lambda_{q,m,p,s} &= \lambda_{m,p,s} + \frac{q+1/2}{2 \pi n \sqrt{r_0 R}} \lambda_{m,p,s}^2, \label{eq:resonatorLambda}
\end{align}
where $\lambda_{m,p,s}$ is the cutoff wavelength that bounds the microresonator profile, $\zeta_{1,2,3,\cdots} = 2.338,4.088,5.521,\cdots$ are the absolute values of the Airy function roots, which can be approximated $\zeta_p \approx \left(\frac{3}{8} \pi (4p-1) \right)^{2/3}$, $s$=+1/-1 corresponds to TE/TM polarization, $n=1.46$ is the microresonator refractive index, and $r = 15\, \mu$m is the physical radius of the fiber. \eqref{eq:resonatorLambda} describes the resonance wavelengths for an ideal parabolic microresonator with radius of curvature $R$. Approximate azimuthal and axial separations, using nominal and not precisely measured radius and index values, are
\begin{align}
\lambda_{97,1,s} -\lambda_{98,1,s} &= 17\text{ nm} \\ 
\lambda_{97,1,s} -\lambda_{97,2,s} &= 92\text{ nm}.
\end{align}

The axial mode series is bounded from above by a cutoff wavelength that is directly proportional to the ERV through the rescaling relation
\begin{equation}\label{eq:rescaling}
\frac{\Delta r}{r_0} = \frac{\Delta \lambda}{\lambda_0},
\end{equation}
where $\Delta r = r(z) - r_0$ is the variation of the radius at position $z$ from the reference radius $r_0$, $\lambda_0$ is the cutoff wavelength for radius $r_0$, $\Delta \lambda = \lambda - \lambda_0$, and the relation is valid for small radius variation where $\Delta r \ll r_0$. This method interrogates the radius profile $r(z)$ of the TUT with subangstrom ERV precision. Characterization of taper profiles was performed by Birks, Knight, and Dimmick, who found small variation in the radial profile of tapers with nominally 20 $\mu$m diameter, which is negligible for most applications involving light traveling through a taper, but is a substantial change for whispering gallery modes traveling around one \cite{Birks2000}.

The SNAP bottle microresonators reported herein have a 30 $\mu$m nominal waist diameter with a 2 mm long waist region. The tapering process does not give rise to identical microresonators each time, but they consistently have hundreds of axial modes with order 100 nm ERV height, loaded quality factors $Q \approx 10^6$, and 6 GHz frequency separation between axial modes. Spectrograms characterizing two exemplar microresonators, labeled tapers A and B, are shown in Fig.\ \ref{fig:spectrograms}. The taper waist region lies between the two minima that separate this region from the neighboring transition regions. We use the extent of the highest-order axial mode to characterize the longitudinal extent of the microresonator, as this is the distance along which the probe taper excites microresonator modes. Let this microresonator width be $W_M^{(i)}$, where $i\in\{A, B\}$ indexes the taper, $q_{\text{max}}$ be the highest-order bound axial mode, $\Delta \lambda_i = \lambda_{q_{\text{max}},m,p,s}^{(i)} - \lambda_{1,m,p,s}^{(i)}$ be the axial mode bandwidth, and $\Delta \Lambda_i = \lambda_{q,m+1,p,s}^{(i)} - \lambda_{q,m,p,s}^{(i)}$ be the azimuthal mode spacing. Microresonators that are better centered within the waist have larger extent and ERV height.

\begin{table*}[t]
\caption{Microresonator properties. Fixed fit parameter $x_0$ and free fit parameters $h$, $L$, and $R$, for microresonator profiles, in addition to the axial bandwidth $\Delta f$ (the frequency-space counterpart of $\Delta \lambda$), the height of the microresonator in effective radius (ERV) space [see \eqref{eq:rescaling}] $\overline{\Delta r}$, the number of axial modes $N$, the microresonator spatial width $W_m$, and the average spacings between axial modes $\overline{\Delta \nu}$ (over a full axial series) and $\overline{\Delta \nu}_r$ (over the reduced regions with superior uniformity indicated in Fig.\ \ref{fig:modeSpacing}). }
\label{tab:fitParameters}
\begin{center}
\begin{tabular}{|c||c|c|c|c||c|c|c|c|c|c|} \hline
 Taper & $x_0$ [$\mu$m] & $h$ [nm] & $L$ [nm/$\mu$m$^2$] & $R$ [nm/$\mu$m$^2$] & $\Delta f$ [THz] &  $\overline{\Delta r}$ [nm] & $N$ & $W_m$ [mm] & $\overline{\Delta \nu}$ [GHz] & $\overline{\Delta \nu}_r [GHz]$ \\ \hline 
 A & 1413 & 1548.2 & $-2.42 \times 10^{-5} $ & $-3.05 \times 10^{-6}$ & 1.3 & 96 & 213 & 1.4 & $6.4 \pm 0.6$ & $6.5 \pm 0.2$ \\ \hline  
 B & 2324 & 1563.4 & $-2.37\times 10^{-5}$ & $-3.56 \times 10^{-6}$ & 2.4  & 184 & 412 & 1.9 & $5.8 \pm 0.4$ & $6.0 \pm 0.2$ \\
 \hline
\end{tabular}
\end{center}
\end{table*}

For frequency comb generation, it is desirable to make a microresonator with modes that exhibit equidistant eigenfrequency spacing over a broad bandwidth. Equidistant spacing optimizes phase matching and thus the efficiency of nonlinear processes driving comb generation, while broad bandwidth, in addition to being a figure of merit in its own right, enables self-referencing and synthesis of pulses with shorter duration. Microresonators made through tapering are observed to be more than ten times larger in ERV than can be made through the standard laser-induction method based on annealing the fiber, which is limited by the tension frozen into the fiber upon fabrication \cite{Sumetsky2011}, and support a larger number of axial modes across a broader bandwidth. The bandwidth spanned by the modes of these microresonators could be expanded by setting the azimuthal mode spacing such that equidistant frequency spacing is maintained across axial series corresponding to modes with different azimuthal quantum numbers $m$, as discussed in \cite{Dvoyrin2016}. The minimal condition for this is that the azimuthal spacing $\Delta \Lambda$ is smaller than the full bandwidth of the series of axial modes. The microresonator on taper B satisfies this criterion, though fine control through temperature or stress tuning may be necessary to optimize the eigenfrequency matching and mode coupling conditions.

We find that the profiles of the fabricated microresonators are well fit by a function of the form
\begin{equation} \label{eq:biparabola}
f(x) = h + L \times (x-x_0)^2 H(x+x_0)  + R \times (x-x_0)^2 H(x-x_0),
\end{equation}
where $h$ is the height of the microresonator in wavelength space positioned with its maxima at $x_0$, $H(x-x_0)$ is the Heaviside step function, $L$ and $R$ are fit parameters characterizing the parabolic curvature for the left- and right-hand sides respectively, and this assumes $q \gg1$. Microresonators with profiles of this form have equidistant modal frequency spacing even if $L \neq R$, as long as the profile is continuous \cite{Sumetsky2015}. Fits are shown in Figs.\ \ref{fig:spectrograms}(c) \& \ref{fig:spectrograms}(f), with best fit parameters given in Table \ref{tab:fitParameters}. This $L \neq R$ asymmetry offers a hint in the ongoing investigation into what processes shape these microresonators.

The frequency spacing between adjacent microresonator axial modes $\Delta \nu(q) = \nu_{q+1} -\nu_{q}$ are directly assessed in Fig.\ \ref{fig:modeSpacing}, and can be quantified with the relative standard deviation over the full axial mode range, which is 9\% and 7\% for tapers A and B respectively. The dashed lines enclose regions with superior uniformity and relative standard deviations of 3\%  and 4\% for tapers A and B respectively. A frequency comb made using a microresonator with azimuthal spacing such that the modes at the edges of the uniform region overlap would preferentially transfer power to the equally spaced modes and offer superior uniformity compared to one that used the full axial range. We suggest that this uniformity may be improved with further optimization of the tapering process.

In summary, we demonstrate that the simple procedure of tapering fiber to $\sim 30 \, \mu$m waist diameter gives rise to SNAP bottle microresonators in the resulting waist region that have large effective radius variation with hundreds of axial modes, loaded $Q \approx 10^6$, and a nearly harmonic profile that gives rise to good uniformity in eigenfrequency spacing. Refinement of the fabrication process towards precision control that optimizes frequency equidistance and matching of adjacent azimuthal orders would yield a promising method for fabricating frequency comb generators and delay lines.

\section*{Funding and Acknowledgements}
M.S. acknowledges the Royal Society Wolfson Research Merit Award (WM130110). The authors acknowledge funding from the Horizon 2020 Framework Programme (H2020) (H2020-EU.1.3.3, 691011), Engineering and Physical Sciences Research Council (EPSRC) (EP/P006183/1), and US Army Research Laboratory (ARL) (W911NF-17-2-0048). Portions of this work were presented at SPIE Photonics Europe 2018, \cite{Gardosi2018} and at CLEO 2018 \cite{Vitullo2018}.

\bibliography{snapVIAtapering}


\end{document}